\documentclass[twocolumn,secnumarabic,amssymb, nobibnotes, aps, prl]{revtex4-1}

\usepackage{graphicx}
\usepackage{gensymb}
\usepackage{color}
\usepackage[utf8]{inputenc}
\usepackage[T1]{fontenc}

\begin{document}

\title{Thermal transport in 2D and 3D nanowire networks}%

\author{Maxime Verdier$^{1}$}
\author{David Lacroix$^{1}$}
\author{Konstantinos Termentzidis$^{2}$}
\email{Konstantinos.Termentzidis@insa-lyon.fr}

\affiliation{
$^{1}$ Universit$\acute{e}$ de Lorraine, LEMTA UMR 7563, 54505 Vandoeuvre les Nancy, France\\
$^{2}$ Univ Lyon, CNRS, INSA-Lyon, Universit$\acute{e}$ Claude Bernard Lyon 1, CETHIL UMR5008, F-69621, Villeurbanne, France}

\date{2 February, 2018}%

\begin{abstract}
We report on thermal transport properties in 2 and 3 dimensions interconnected nanowire networks (strings and nodes). The thermal conductivity of these nanostructures decreases in increasing the distance of the nodes, reaching ultra-low values. This effect is much more pronounced in 3D networks due to increased porosity, surface to volume ratio and the enhanced backscattering at 3D nodes compared to 2D nodes. We propose a model to estimate the thermal resistance related to the 2D and 3D interconnections in order to provide an analytic description of thermal conductivity of such nanowire networks; the latter is in good agreement with Molecular Dynamic results.       
\end{abstract}

\maketitle

New innovating and hightly sophisticated architectured nanostructures are now feasible with the rapid evolution of the elaboration methods~\cite{rauber_highly-ordered_2011,li_fabrication_2017,galland_fabrication_2013,han_bio-inspired_2014}. Among them 2D and 3D networks of nanowires are a new class of nanostructured materials with interesting mechanical, optical, electronic and thermal properties. 2D networks are proposed as optoelectronic or biological devices and sensors due to their mechanical strength and flexibility~\cite{han_bio-inspired_2014}. Furthermore, 2D or 3D ordered or disordered networks could be useful for complex integrated nanoelectronic circuits~\cite{romo-herrera_covalent_2007}. Independently of their application, their main characteristics are the extremely low mass density, the high surface to volume ratio as well as high porosity and their remarkable mechanical properties. In the literature, 3D networks have been elaborated during the last decade at the nanoscale with several different materials (silver~\cite{madaria_uniform_2010}, manganese dioxide~\cite{jiang_ultrafine_2011} or silicon~\cite{mulazimoglu_silicon_2013,ge_orientation-controlled_2005}).

There are three main fields of applications for silicon nanowire (NW) networks and nanomeshes. (i) Thermoelectricity (TE): The huge porosity and surface-to-volume ratio of such nanostructures reduce strongly their lattice thermal conductivity (TC), making Si NW networks promising candidate for TE applications \cite{yu_reduction_2010,ravichandran_coherent_2014,hochbaum_enhanced_2008,boukai_silicon_2008}. (ii) Transistors: These systems can be easily integrated in nanoelectronic devices (Si compatible) and could be the next generation of transistors thanks to their high density of nanowire interconnections \cite{hochbaum_controlled_2005,heo_large-scale_2008,wang_general_2004}. (iii) Catalysis: Nanowire networks are interesting for catalysis applications because of their large surface-to-volume ratio that allows improved efficiency of chemical reactions. Furthermore, their strong mechanical robustness as compared to isolated nanowires or nanoparticles make them interesting candidates to practically achieved all these innovative applications \cite{rauber_highly-ordered_2011,wei_geometric_2016}.

Concerning the thermal properties of nanostructures, they have attracted high attention for various applications in the fields of microelectronics, optoelectronics and energy harvesting. Nanoscale heat transfer is known to diverge from classical physics \cite{cahill_nanoscale_2014}, especially in semi-conductors where heat is mostly carried by lattice vibrations (phonons). Interestingly, nano-structuration usually reduces the TC due to boundary scattering while the electrical properties could be preserved \cite{lee_nanoporous_2008}. The design of nanostrucured materials with ultra low TC beating sometimes the amorphous limit while keeping large crystalline fraction is now possible~\cite{mizuno_beating_2015,verdier_crystalline-amorphous_2016}.

In this work, we focus on a specific architectured nanostructure which consists of interconnected nanowires with square cross-section forming a 2D or 3D network. The 2D networks (or nanomeshes) have been studied in the last decade mainly due to their low TC but also as nanostructures in which coherent effects might be observed~\cite{yu_reduction_2010,lee_investigation_2017}. Contrarily to 2D networks, thermal properties of 3D NW networks have not been investigated yet. In this work, a systematic study of their heat transport properties, depending on their geometry and their dimensionality (2D or 3D), is conducted by means of Molecular Dynamics (MD) simulations. For the 2D networks, the TC is computed in the in-plane direction. The nanowires have a square cross section with dimensions $d \times d$ and they are interconnected with 90\degree angle (fig. \ref{fig:networks}). Details of the simulation methodology are given in Appendix A.

\begin{figure}[!htb]
	\centering
	\includegraphics[angle=0,width=0.9\linewidth]{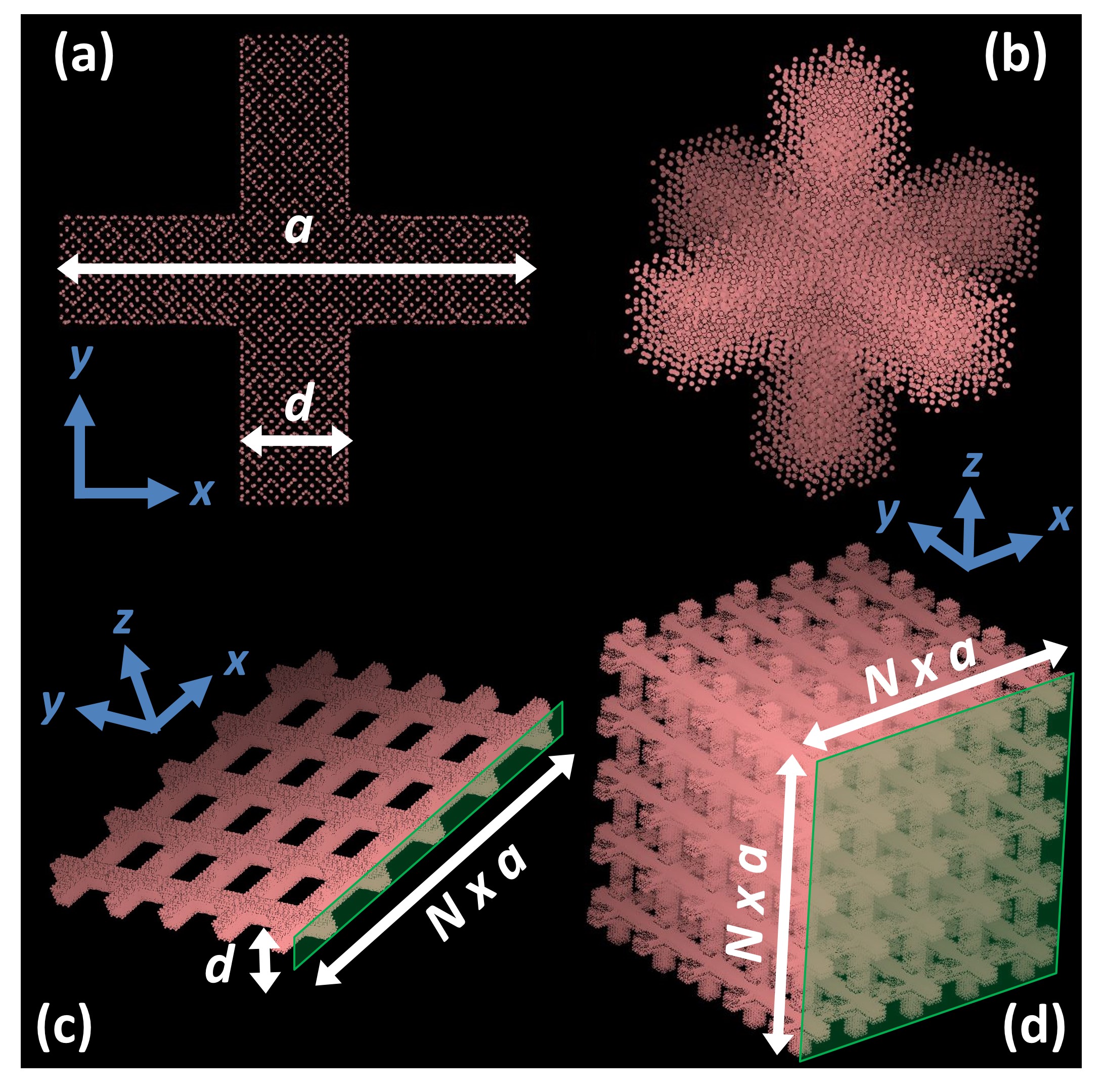}
	\caption{\footnotesize Visualization of Molecular Dynamics systems of 2D and 3D nanowire networks. Simulation cell containing one node for (a) 2D and (b) 3D networks. (c) and (d): Representation of global modeled systems thanks to periodic boundary conditions. The green surfaces represent the total cross section of each system when considering the TC in the direction perpendicular to these surfaces.}\label{fig:networks}
\end{figure}

\begin{figure}[!htb]
	\centering
	\includegraphics[angle=0,width=0.8\linewidth]{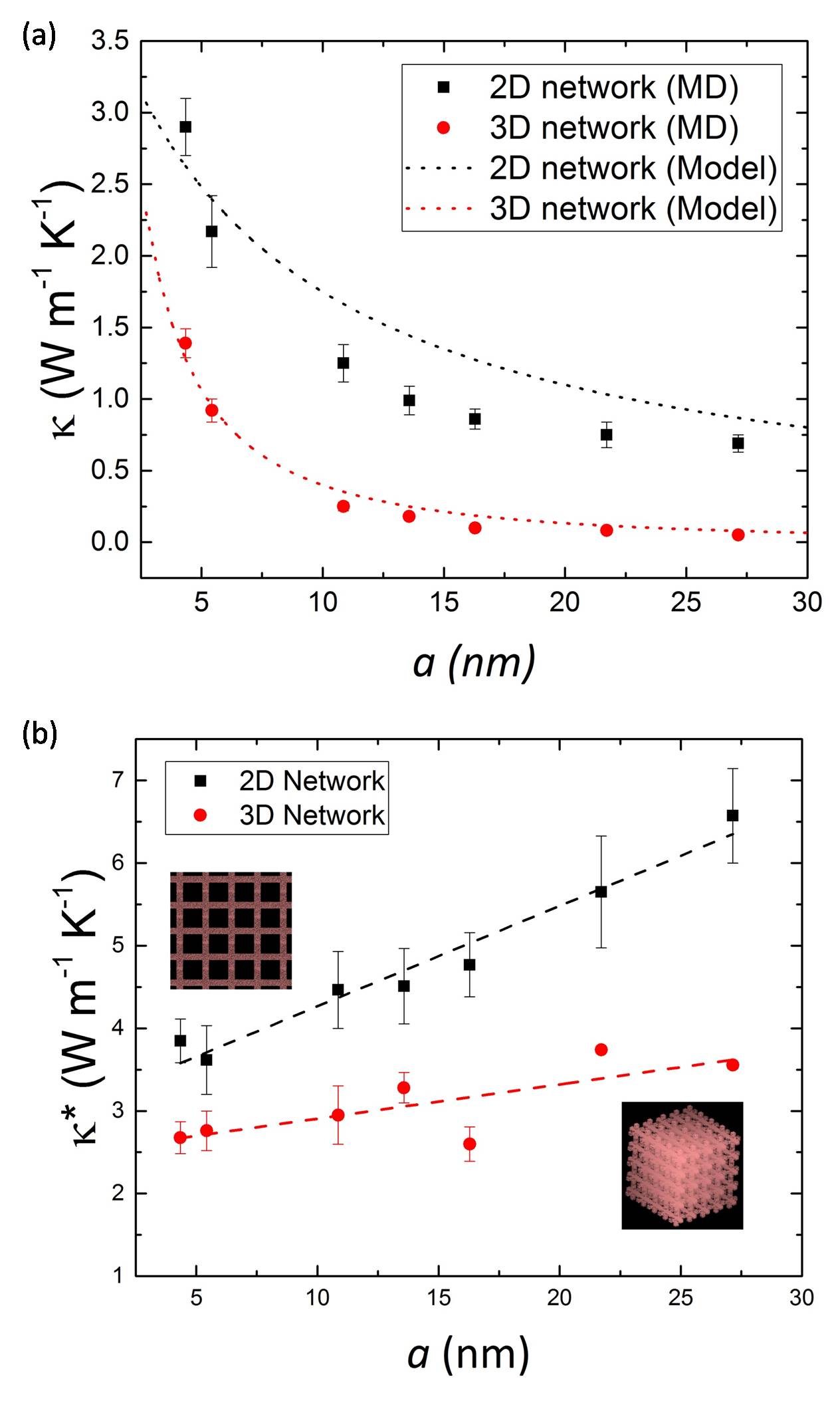}
	\caption{\footnotesize Thermal conductivity of 2D and 3D nanowire networks as a function of period with a constant nanowire diameter $d=2.715$~nm at room temperature. (a) Thermal conductivity $\kappa$ obtained from EMD simulations, comparison with the thermal resistance model (eq. \ref{eq:ktot_2D} and \ref{eq:ktot_3D}). (b) Thermal conductivity corresponding to an equivalent non-porous medium ($\kappa^*=\kappa(1+\phi)/(1-\phi)$).}\label{fig:results_period}
\end{figure}

First, the effect of the period $a$ (the distance between two nodes centers) on thermal conduction is investigated. The cross section of the nanowires is set to $2.715 \times 2.715$~nm ($5~a_0 \times 5~a_0$). The TC $\kappa$ of 2D and 3D nanowire networks is depicted in fig \ref{fig:results_period}a as a function of the period. The TC decreases when the distance between nodes increases. This can be understood in terms of porosity ($\phi$) and surface-to-volume ratio, which both increase upon increasing the period. Phonon boundary scattering occurs more often and thermal transport through the structure is hindered, especially for large periods. The reduction of TC as compared to the bulk ($\kappa_{bulk} \simeq 150$~W~m$^{-1}$~K$^{-1}$ at 300~K \cite{glassbrenner_thermal_1964,maycock_thermal_1967}) can reach three or four orders of magnitude in 2D and 3D networks, respectively. This means that TC of these structures can be well below the amorphous limit ($\kappa_{amorphous} \simeq 1.5$~W~m$^{-1}$~K$^{-1}$ at 300~K \cite{wada_thermal_1996}). Such low values seem surprising but can be explained by the extremely high porosity, which reaches 81\% for 2D networks and 97\% for 3D networks studied here. Moreover, the $S/V$ ratio is very large and phonon mean free path is drastically reduced \cite{jain_phonon_2013,jean_monte_2014}. All networks have a lower TC than a single nanowire of same cross section $d \times d$, for which $\kappa$ is found to be about 11~W~m$^{-1}$~K$^{-1}$ with MD.

In order to distinguish the effects of porosity and nano-structuration on thermal transport, the effective thermal conductivity $\kappa^*$ for an equivalent non-porous medium has been computed for the systems with $d=2.715$~nm and different periods. The TC obtained with MD simulations can be written as
\begin{equation}
\kappa = \kappa_{bulk} f^* f(\phi)
\end{equation}
with $f(\phi)$ the correction factor representing the reduction of the TC due to the porosity, and $f^*$ the factor accounting for nano-structuration effects (phonon backscattering at free surfaces, coherent effects, etc) that depends on several parameters as the $S/V$ ratio. The effective TC is defined as $\kappa^*=\kappa_{bulk} f^*$. Thus, the effect of the porosity does not appear in the effective TC and the reduction of $\kappa^*$ is only due to the nanostructuration. $f(\phi)$ is taken from the Maxwell-Garnett Effective Medium Model (EMM) \cite{lee_investigation_2017}:
\begin{equation}
f(\phi)=(1-\phi)/(1+\phi)
\end{equation}
Thus, the effective TC can be calculated from the value given by Molecular Dynamics with $\kappa^*=\kappa/f(\phi)=\kappa(1+\phi)/(1-\phi)$.

The effective TC goes from 3.8 to 6.6~W~m$^{-1}$~K$^{-1}$ for 2D networks and 2.7 to 3.6~W~m$^{-1}$~K$^{-1}$ for 3D networks as $a$ goes from 4 to 27~nm (fig. \ref{fig:results_period}b), which seem reasonable given the huge $S/V$ ratios. In contrast with the behavior of the TC $\kappa$, it is found for both 2D and 3D systems that $\kappa^*$ slightly increases when the period increases, as expected from ref \cite{lee_investigation_2017} due to less impact of backscattering at the nodes. Thus, the decrease of $\kappa$ when increasing the period is mainly due to the growing porosity.

In spite of the increasing $S/V$ ratio which should lead to more phonon scattering, $\kappa^*$ increases with the period~\cite{lee_investigation_2017}. When considering one direction of measurement (direction of the heat flux), phonon scattering on the nanowires walls are not always resistive to heat transport. In nanowires parallel to the direction of interest, if scattering is fully diffuse, there is only 50\% chance for each scattered phonon to go back (backscattering); while at the crossings with the perpendicular nanowires, which do not contribute to heat transport in the direction of measurement, phonons colliding with the walls are necessarily scattered backward. Moreover, it has been shown that free surfaces modeled in Molecular Dynamics have a great specularity, even at room temperature \cite{verdier_hierarchical_nodate}. In the case of fully specular walls, nanowires parallel to the heat flux are supposed not to be resistive at all, while perpendicular nanowires would lead to 100\% of back-scattering. Thus, the nodes hinder the TC more than the intrinsic thermal resistance of the nanowires. Increasing the period, the nodes move away from each others and there is less resistance to thermal transport, even if there is more scattering surface in the system. Thus, $\kappa^*$ increases with the period. However, for very long periods, not considered here, the effective TC shall reach saturation to the TC of a single nanowire with cross section $d \times d$ ($\sim 11$~W~m$^{-1}$~K$^{-1}$). For 3D networks, there are more phonon reflections at nodes than in 2D networks, this explains the lower effective TC in 3D networks.

\begin{figure}[!htb]
	\centering
	\includegraphics[angle=0,width=0.9\linewidth]{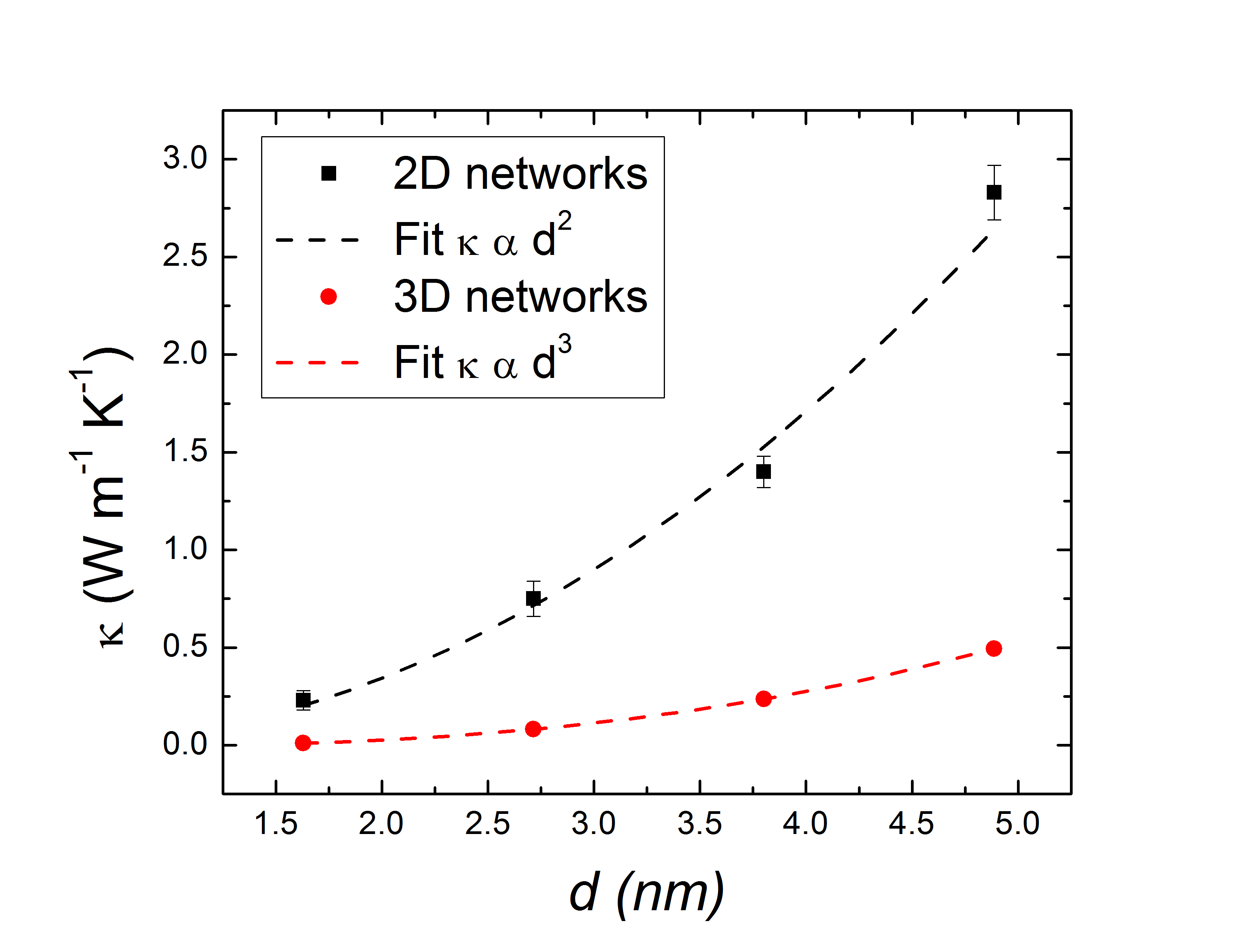}
	\caption{\footnotesize Thermal conductivity of 2D and 3D nanowire networks as a function of nanowire diameter for a constant period $a=21.72$~nm at $T=300$~K.}\label{fig:results_diameter}
\end{figure}

The impact of the diameter of the nanowires on thermal transport has also been investigated. In fig. \ref{fig:results_diameter} is depicted the TC $\kappa$ of 2D and 3D nanowire networks with a constant period $a=21.72$~nm as a function of the nanowires dimension $d$. Obviously, increasing the cross section of the nanowires, thermal transport is enhanced. Decreasing the diameter, the TC of 2D networks drops below the amorphous limit while porosity varies from 60 to 85\%. For 3D structures, $\kappa$ is less than 1~W~m$^{-1}$~K$^{-1}$ for all diameters, and the porosity is between 87 and 98\%. For $d \simeq 5$~nm, the TC of the 3D network is divided by 300 as compared to the bulk. These dimensions can already be reached with current fabrication methods for MnO$_2$ nanowire networks \cite{jiang_ultrafine_2011}. Finally, we notice that the diameter dependent TC of 2D networks increases with $d^2$ while 3D networks follow a $d^3$ law. This observation is valid when $d$ is small compared to the period. When $d$ tends toward $a$, the TC of 2D and 3D systems are expected to reach the values of a nanofilm of thickness $d$ ($\kappa \simeq 23$~W~m$^{-1}$~K$^{-1}$) and of bulk silicon ($\kappa \simeq 160 $~W~m$^{-1}$~K$^{-1}$), respectively.

Interestingly, heat conduction is always hindered further in 3D network than in 2D network, even for a same $S/V$ ratio (TC as a function of $S/V$ is plotted in Appendix B). For example, for $S/V \simeq 1.41$~nm$^{-1}$ the TC is about 0.8~W~m$^{-1}$~K$^{-1}$ in 2D networks and less than 0.1~W~m$^{-1}$~K$^{-1}$ in 3D networks). This observation is even valid when considering $\kappa^*$ ($\kappa^*_{2D}\simeq$4.8~W~m$^{-1}$~K$^{-1}$ and $\kappa^*_{3D}\simeq$3.7~W~m$^{-1}$~K$^{-1}$ at the same $S/V \simeq 1.41$~nm$^{-1}$). This phenomenon is counter-intuitive, as reduction of the dimensionality usually leads to lower TC.

To explain the lower thermal transport in 3D structures, a model based on the use of thermal resistances has been developed. The derivation of this model is given in Appendix C to this work and results in the following expressions for the TC: 
\begin{equation}
\kappa^{2D}=\frac{1}{d \left( R_{NW}+R_{node}^{2D} \right)}
\end{equation}\label{eq:ktot_2D}
\begin{equation}
\kappa^{3D}=\frac{1}{a \left( R_{NW}+R_{node}^{3D} \right)}
\end{equation}\label{eq:ktot_3D}
with $R_{NW}$ the thermal resistance of a portion of nanowire between two nodes computed from EMD simulations. $R_{node}$ is the thermal resistance of a node. It is chosen as the adjustable parameter and it is different for 2D and 3D networks because the number of interconnected nanowires is not the same.

In fig. \ref{fig:results_period} the model is compared to EMD results for 2D and 3D nanowire network with constant size of nanowire $d=2.715$~nm and varying period. Simulations and model are in a particularly good quantitative agreement for 3D networks. The model correctly reproduces the trend for both 2D and 3D networks and predicts that thermal conduction in 3D systems is always lower than in 2D systems. This phenomenon mainly comes from the difference in total cross sections of 2D and 3D networks, which leads to a factor $a/d$ between $\kappa^{2D}$ and $\kappa^{3D}$ in the model, assuming that $R_{node}^{2D} \simeq R_{node}^{3D}$. This factor is approximately retrieved from EMD results.

The best fit was obtained with $R_{node}=1.2 \times 10^8$~K~W$^{-1}$ for 2D network and $R_{node}=1.6 \times 10^8$~K~W$^{-1}$ for 3D network. For the sake of comparison, $R_{NW}$ varies from $0.2\times 10^8$ to $3.0 \times 10^8$~K~W$^{-1}$ as $a$ goes from 4 to 27~nm. The thermal resistance of a node is roughly equivalent to a 15~nm portion of nanowire with $d=2.715$~nm, whereas the length of a node is only $d$. The impact of the nodes on thermal conduction is huge and cannot be neglected. At each node, some phonons experience back scattering because of the free surface of perpendicular nanowires and this greatly reduces thermal transport \cite{lee_investigation_2017,verdier_heat_2017}. Moreover, the node resistance is found to be slightly higher for 3D networks than 2D networks. This is related to the fact that 3D nodes have 4 perpendicular branches (compared to 2 perpendicular branches for 2D nodes), so more backscattering occurs at 3D nodes. This also confirms what has been claimed above explaining why even $\kappa^*$ is lower for 3D systems than for 2D ones: this is due to the more important scattering at 3D nodes.

To conclude, we investigated thermal conduction in 2D and 3D networks of interconnected Silicon nanowires and we showed that the TC is drastically reduced in such structures due to a combination of large porosity and increased backscattering at the nodes. The lowering of thermal transport is more pronounced in the 3D networks than in the 2D networks because 3D structures have higher porosity and their nodes have more branches which lead to increased backscattering and larger thermal resistance. A model based on equivalent thermal resistances reproduces the main trends of the MD results and confirms these interpretations. The small discrepancy between model and simulations could arise from correlations between resistances, or from new vibrating modes emerging for specific dimensions of the networks.

\section*{Acknowledgements}

Calculations were performed on the EXPLOR Mesocenter (University of Lorraine).

\section*{Appendix A: Simulation methodology}

All simulations were performed with LAMMPS open source software \cite{plimpton_fast_1995}, using the Stillinger-Weber potential for silicon \cite{stillinger_computer_1985} with modified coefficients \cite{vink_fitting_2001}. The structures are built from a slab of bulk crystalline silicon deleting atoms of certain regions to obtain one ``node'' (figs. 1a and 1b of the main article) of the nanowire's network. Periodic boundary conditions are applied along two or three directions to model an infinite 2D or 3D nanowire networks, respectively (figs. 1c and 1d of the main article). Then a conjugate gradient minimization is done and the structures are relaxed at 300~K under NVT ensemble during 200~ps. Finally, the thermal conductivity at room temperature is extracted thanks to Green-Kubo formalism, estimating the correlation of flux fluctuations during 10~ns with a time window of 40~ps. Computational details can be found in previous works \cite{verdier_crystalline-amorphous_2016,verdier_heat_2017}. Size effects due to the small size of the simulation box were checked, modeling a bigger structure containing four nodes of a 2D system. The difference between computed thermal conductivities with one and four nodes is less than 7\% and remains within the error bars.

\section*{Appendix B: Thermal conductivity as a function of the surface-to-volume ratio}

\begin{figure}[!htb]
	\centering
	\includegraphics[angle=0,width=0.8\linewidth]{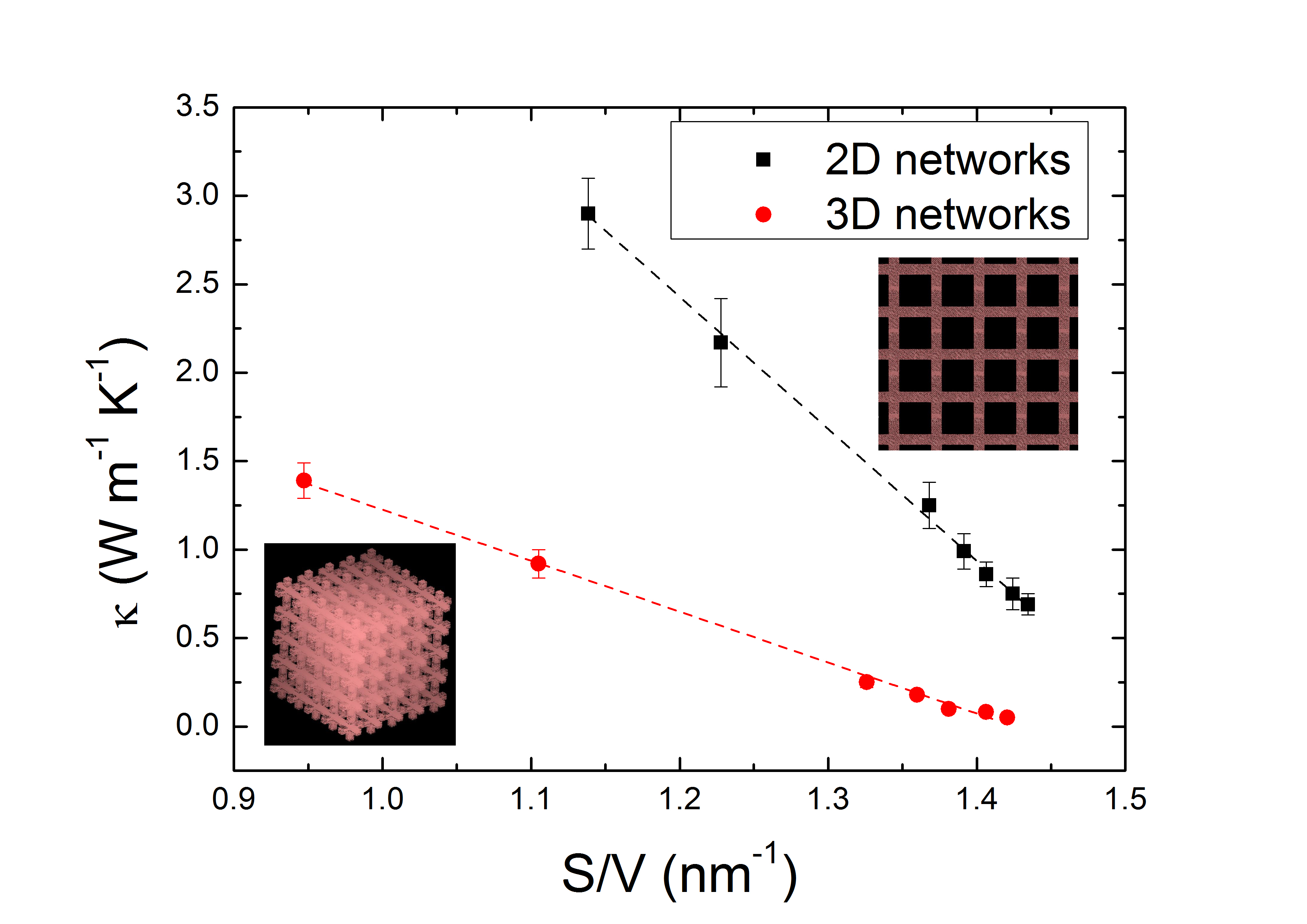}
	\caption{\footnotesize Thermal conductivity of 2D and 3D nanowire networks as a function of the surface-to-volume ratio $S/V$ for a constant nanowire diameter $d=2.715$~nm at $T=300$~K. The thermal conductivities in this figure are those presented in fig. 2a of the main article. The dashed lines are linear fits.}\label{fig:k_vs_SV}
\end{figure}

In nanostructures, heat transport is usually controlled by the surface-to-volume ratio $S/V$, which is inversely proportional to the phonon boundary mean free path linked to the nanostructuration ($\Lambda \simeq 4V/S$). In the structures studied in this work, the boundary mean free path is between 2.7~nm (for $S/V \simeq 0.95$~nm$^{-1}$) and 4.2~nm (for $S/V \simeq 1.45$~nm$^{-1}$). When $S/V$ increases, the thermal conductivity decreases due to enhanced scattering on the free surfaces. In figure \ref{fig:k_vs_SV} is plotted the thermal conductivity of 2D and 3D nanowire networks obtained with Molecular Dynamics as a function of $S/V$. For a same $S/V$, the thermal conductivity of 3D networks is always lower than that of 2D networks. This means that the $S/V$ ratio is not sufficient to describe the reduction of thermal transport in nanostructures with different geometries. The reason is that the $S/V$ ratio does not take into account the surface orientation which influences the degree of phonon backscattering. For both 2D and 3D networks, the thermal conductivity follows a linear trend given by
\begin{equation}
\kappa^{2D} = 11.4 - 7.5 S/V
\end{equation}
\begin{equation}
\kappa^{3D} = 4.1 - 2.9 S/V
\end{equation}

\section*{Appendix C: Thermal resistances model}

\begin{figure}[!htb]
	\centering
	\includegraphics[angle=0,width=\linewidth]{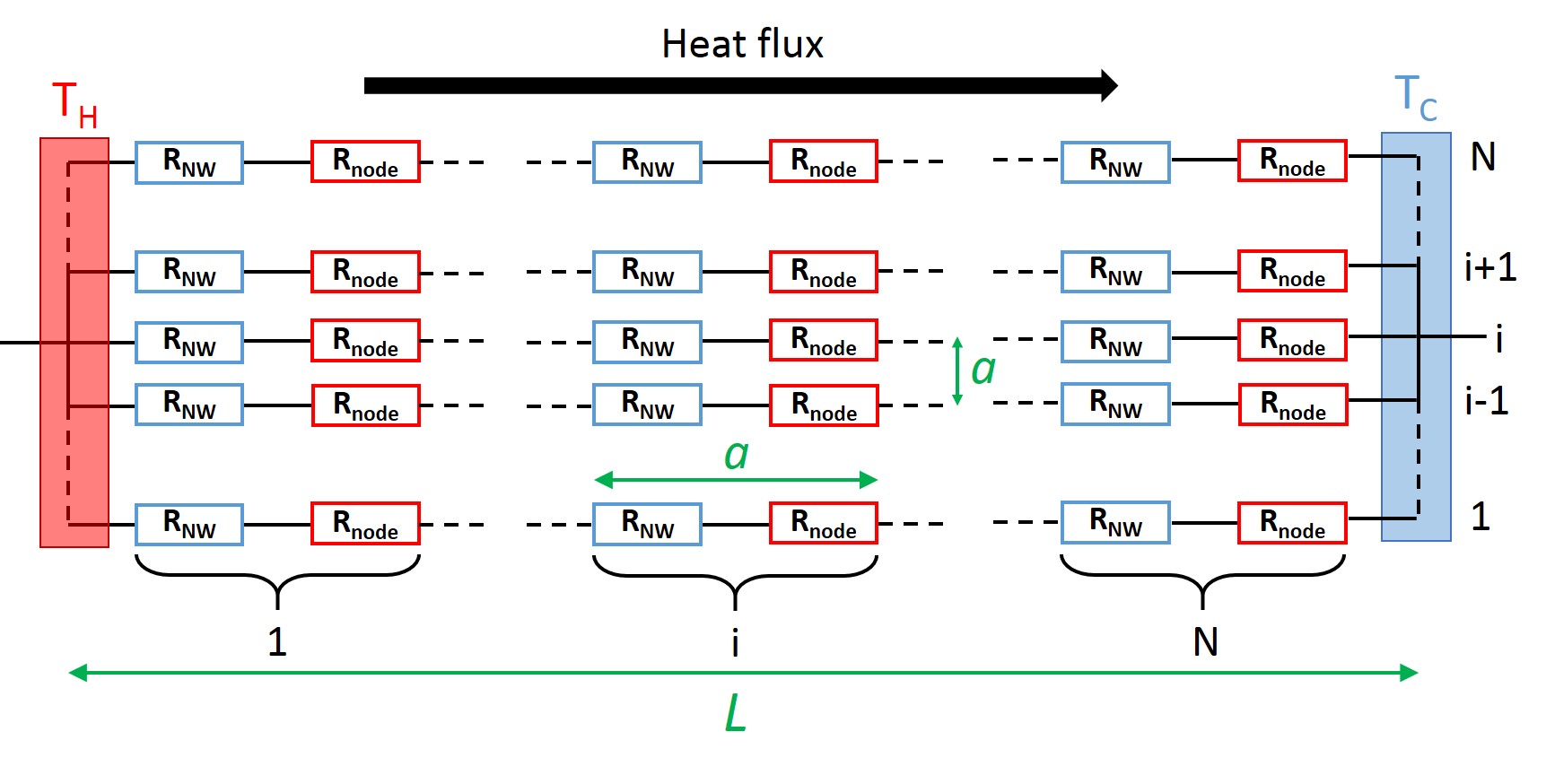}
	\caption{\footnotesize Schematic of the thermal resistances model for a 2D nanowire network with $N \times N$ periods.}\label{fig:thermal_resistances}
\end{figure}

When a temperature gradient is applied within the network nanostructure, heat carriers are subject to two types of thermal resistances (fig. \ref{fig:thermal_resistances}). The first one is $R_{NW}$ the thermal resistance of a short nanowire (``strut'') between two nodes (blue in fig. \ref{fig:thermal_resistances}), which depends on the cross section $d \times d$ and the length $a-d$ of each portion of nanowire (see eq. 10\ref{eq:R_NW}). The second one is the thermal resistance related to the nodes $R_{node}$ (red in fig. \ref{fig:thermal_resistances}), which is unknown. By analogy with macroscopic heat transfer, the heat flux per surface area $J$ in the direction of the temperature gradient is given by Fourier's law
\begin{equation}
J=- \kappa \frac{T_H-T_C}{L}
\end{equation}\label{eq:fourier}
with $T_H$ and $T_C$ the temperatures of hot and cold thermostats, respectively, and $L=N \times a$ the distance between the two thermostats (see fig. \ref{fig:thermal_resistances}). To reproduce the results of the present work, $N$ has to tend toward infinity. Thus, the model describes a cubic (or square) infinite nanowire network.

With respect to the thermal resistance formalism, the heat flux can also be written as
\begin{equation}
J=\frac{T_H-T_C}{R_{tot}S}
\end{equation}\label{eq:flux_resistance}
where $R_{tot}$ is the total thermal resistance and $S$ is the total cross section (perpendicular to the flux) of the system. The total cross section of 2D networks is $S=Na \times d$, whereas for 3D network $S=Na \times Na$ (see fig. 1 in the main article). From equations 3\ref{eq:fourier} and 4\ref{eq:flux_resistance}, we derive the expression for thermal conductivity:
\begin{equation}
\kappa=\frac{L}{R_{tot}S}
\end{equation}\label{eq:k}
in which the total thermal resistance is different for 2D and 3D networks:
\begin{equation}
\frac{1}{R_{tot}^{2D}}=\frac{1}{R_{NW}+R_{node}^{2D}}
\end{equation}\label{eq:Rtot_2D}
\begin{equation}
\frac{1}{R_{tot}^{3D}}=\frac{N}{R_{NW}+R_{node}^{3D}}
\end{equation}\label{eq:Rtot_3D}
Combining the last three equations and replacing \mbox{$L=Na$} and $S=Na \times d$ (2D) or $S=Na \times Na$ (3D), it comes two simple expressions for thermal conductivity:
\begin{equation}
\kappa^{2D}=\frac{1}{d \left( R_{NW}+R_{node}^{2D} \right)}
\end{equation}\label{eq:ktot_2D}
\begin{equation}
\kappa^{3D}=\frac{1}{a \left( R_{NW}+R_{node}^{3D} \right)}
\end{equation}\label{eq:ktot_3D}

In order to determine $R_{NW}$, the thermal conductivity of an infinitely long nanowire with cross section $2.715 \times 2.715$~nm has been computed with EMD. We obtained $\kappa_{NW}=11 \pm 2$~W~m$^{-1}$~K$^{-1}$. Then the thermal resistance of the portion of nanowire between two nodes is deduced for each system with
\begin{equation}
R_{NW}=\frac{a-d}{\kappa_{NW} d^2}
\end{equation}\label{eq:R_NW}
Finally, $R_{node}$ is chosen as the adjustable parameter. It is considered as a constant parameter for a given $d$, but it is not the same for 2D and 3D networks because the number of interconnected nanowires at a node is different (4 and 6, respectively). Each node corresponds to an abrupt change of cross section of the material for a length $d$. This surely affects thermal transport, but in a manner which is unknown. The proposed model allows to quantify the thermal resistance due to the nodes.

\bibliographystyle{unsrt}
\bibliography{biblio}

\end{document}